\documentclass[11pt,letterpaper]{article}
\usepackage{jheppub}

\newcommand{\CF}{{\cal F}}

\newcommand{\CV}{{\cal V}}

\newcommand{\bx}{{\bf x}}

\newcommand{\diff}{{\rm Diff}}

\newcommand{\p}{\partial}

\newcommand{\be}{\begin{equation}}
\newcommand{\ee}{\end{equation}}
\newcommand{\bea}{\begin{eqnarray}}
\newcommand{\eea}{\end{eqnarray}}

\title{General Covariance in Gravity at a Lifshitz Point}
\author{Petr Ho\v{r}ava}
\affiliation{Berkeley Center for Theoretical Physics and Department of 
Physics\\
University of California, Berkeley, CA, 94720-7300, USA\medskip\\ 
Theoretical Physics Group, Lawrence Berkeley National Laboratory\\
Berkeley, CA 94720-8162, USA}
\abstract{\baselineskip=14pt
This paper is based on the invited talks delivered by the author 
at {\it GR 19: the 19th International Conference on General Relativity and 
Gravitation}, Ciudad de M\'exico, M\'exico, July 2010.}  
\makeatletter
\gdef\@fpheader{\mbox{}}
\makeatother
\toccontinuoustrue
\begin{document}
\maketitle
\baselineskip=14pt
\noindent
Gravity may be the one force of nature we are intuitively most familiar with, 
but its theoretical understanding -- despite the beauty of general relativity 
and string theory -- is still shrouded in surprisingly many layers of 
mystery.  Perhaps we already have all the pieces of the puzzle, and just need 
to find the correct way of putting them together; or perhaps new ideas are 
needed.  In this context, the idea of gravity with Lifshitz-type anisotropic 
scaling \cite{mqc,lif,spdim} has attracted a lot of attention recently.  

\vfill\break
In Part~1 of this paper, we briefly review some of the main features of 
quantum gravity with anisotropic scaling, in its original formulation 
initiated in \cite{mqc,lif,spdim}, and comment on its possible relation to 
the causal dynamical triangulations (CDT) approach to lattice quantum 
gravity.  
Part~2 explains the construction of gravity with anisotropic scaling with an 
extended gauge symmetry -- essentially a nonrelativistic version of general 
covariance -- presented in \cite{gen}.  This extra symmetry eliminates the 
scalar graviton polarization, and thus brings the theory closer to general 
relativity at long distances.

\section{Gravity with Anisotropic Scaling}

The central idea of \cite{mqc,lif} is a minimalistic one: to formulate quantum 
gravity as a quantum field theory, with the spacetime metric as the elementary 
field, in the standard path-integral language.  Quantum field theory (QFT) has 
emerged from the 20th century as the universal language for understanding 
systems with many degrees of freedom, ranging from high-energy particle 
physics to condensed matter, statistical physics and more.  Before giving up 
on QFT for quantum gravity, it makes sense to apply its full machinery to this 
problem, without prior restrictions such as microscopic relativistic 
invariance.  The novelty of \cite{mqc,lif} is that gravity is combined with 
the idea of anisotropic scaling of spacetime, more familiar from condensed 
matter, and characterized by
\be
\bx\to b\bx,\qquad t\to b^z t.
\ee
Here $z$ is an important observable, the ``dynamical critical exponent,'' 
associated with a given fixed point of the renormalization group (RG).  
Systems with many different values of $z$ are known, for example in dynamical 
critical phenomena or quantum criticality.  It is natural to ask whether one 
can construct theories with anisotropic scaling and with propagating 
gravitons.  Why?  A consistent theory of gravity with anisotropic scaling can 
be potentially useful for a number of possible applications: 

\begin{itemize}
\item[(i)] Phenomenology of gravity in our Universe of $3+1$ macroscopic 
dimensions.\vspace{-.1in}
\item[(ii)] New gravity duals for field theories in the context of the AdS/CFT 
correspondence; in particular, duals for a broader class of nonrelativistic 
QFTs.\vspace{-.1in}
\item[(iii)] Gravity on worldsheets of strings and worldvolumes of branes.
\vspace{-.1in}
\item [(iv)] Mathematical applications to the theory of the Ricci flow on 
Riemannian manifolds \cite{mqc}.\vspace{-.1in}
\item [(v)] IR fixed points in condensed matter systems, with emergent 
gravitons (new phases of algebraic bose liquids) \cite{eme}.\vspace{-.1in}
\item[(vi)] Relativistic gravity and string theory in asymptotically 
anisotropic spacetimes \cite{aci};
\end{itemize}
and possibly others.  Note that only application (i) is subjected to the 
standard observational tests of gravity, while the others are only constrained 
by their mathematical consistency.  And of course, applications (i--vi) aside, 
this system can serve as a useful theoretical playground for exploring 
field-theory and path-integral methods for quantum gravity.

This approach shares some philosophical background with the idea of asymptotic 
safety, initiated in \cite{weinberg} and experiencing a resurgence of recent 
interest.  Both approaches are equally minimalistic, suggesting that gravity 
can find its UV completion as a quantum field theory of the fluctuating 
spacetime metric, without additional degrees of freedom or a radical departure 
from standard QFT.  While both approaches look for a UV fixed point, 
they differ in the nature of the proposed fixed point:  In asymptotic 
safety, one benefits from maintaining manifest relativistic invariance, and 
pays the price of having to look for a nontrivial, strongly coupled fixed 
point.  In gravity with anisotropic scaling, one gives up Lorentz invariance 
as a fundamental symmetry at short distances, and looks for much simpler, 
perhaps Gaussian or at least weakly coupled fixed points in the UV.  The price 
to pay, if one is interested in application (i), is the need to explain how 
the experimentally extremely well-tested Lorentz symmetry emerges at long 
distances.  

Such Gaussian fixed points of gravity with $z>1$ can also serve as IR fixed 
point in condensed matter systems, as shown for $z=2$ and $z=3$ in \cite{eme}.  
This may be important because it leads to new phases of algebraic bose 
liquids, and gives a new mechanism for making gapless excitations technically 
natural in condensed matter.  Implications for quantum gravity are less 
dramatic:  The gapless excitations at the IR fixed points of \cite{eme} 
are linearized gravitons, only allowed to interact in a way which respects 
linearized diffeomorphism invariance.  Hence, this lattice model is not 
a theory of emergent gravity with nonlinear diffeomorphism symmetries.  

\subsection{The minimal theory}

In our construction of Lifshitz-type gravity, we assume that the spacetime 
manifold $M$ carries the additional structure of a codimension-one foliation 
$\CF$, by $D$-dimensional leaves $\Sigma$ of constant time.  We will use 
coordinate systems $(t,\bx\equiv x^i)$, $i=1,\ldots D$, adapted to $\CF$.  

Perhaps the simplest relevant example of systems with Lifshitz-type anisotropic 
scaling is the Lifshitz scalar theory with $z=2$, 
\be
\label{slifs}
S=\frac{1}{2}\int dt\,d^D\bx\left\{\dot\phi^2-(\Delta\phi)^2\right\},
\ee
with $\Delta$ the spatial Laplacian.  Compared to the relativistic scalar 
in the same spacetime dimension, the Lifshitz scalar has an improved UV 
behavior.  The scaling dimension of $\phi$ changes to $[\phi]=(D-2)/2$, 
and conseqently the (lower) critical dimension also shifts, from the 
relativistic $1+1$, to $2+1$ when $z=2$.  

The most ``primitive'' theory of gravity similar to (\ref{slifs}) would 
describe the dynamics of the spatial metric $g_{ij}(\bx,t)$, invariant under 
time-independent spatial diffeomorphisms.  Because of the lack of 
(time-dependent) gauge invariance, this model would propagate not only 
the tensor polarizations of the graviton, but also the vector and the scalar. 
This ``primitive'' theory becomes more interesting when we make it gauge 
invariant under foliation-preserving diffeomorphisms $\diff(M,\CF)$, 
generated by 
\be
\label{difgen}
\delta t=f(t),\qquad \delta x^i=\xi^i(t,\bx).
\ee
The minimal multiplet of fields now contains, besides $g_{ij}$, also the lapse 
function $N$ and the shift vector $N_i$.  Since the lapse and shift play the 
role of gauge fields of $\diff(M,\CF)$, we can assume that they inherit the 
same dependence on spacetime as the corresponding generators (\ref{difgen}):  
While $N_i(t,\bx)$ is a spacetime field, $N(t)$ is only a function of 
time, constant along $\Sigma$.  Making this assumption about the lapse 
function gives to the minimal theory of gravity with anisotropic scaling, 
sometimes referred to as the ``projectable'' theory \cite{lif}.  (For its 
brief review and some phenomenological applications, see \cite{shinjire}.)   

The dynamics of the projectable theory is described by the most general 
action which respects the $\diff(M,\CF)$ symmetry.  At the lowest orders 
in time derivatives, the action is given by
\be
\label{smint}
S=\frac{2}{\kappa^2}\int dt\,d^D\bx\,\sqrt{g}\,N\left(K_{ij}K^{ij}-\lambda K^2
-\CV\right),
\ee
where 
\be
K_{ij}\equiv\frac{1}{2N}\left(\dot g_{ij}-\nabla_i N_j-\nabla_j N_i\right)
\ee
is the extrinsic curvature of $\Sigma$, $K=g^{ij}K_{ij}$, $\lambda$ is a 
dimensionless coupling, and the potential term $\CV$ is an arbitrary 
$\diff(\Sigma)$-invariant local scalar functional built out of $g_{ij}$, 
its Riemann tensor and the spatial covariant derivatives, but no time 
derivatives.  

Which terms in $\CV$ are relevant will depend on our choice of $z$ at short 
distances.  Terms with $2z$ spatial derivatives have the same classical 
scaling dimension as the kinetic term, and their quadratic part defines the 
Gaussian fixed point.  Terms with fewer derivatives represent relevant 
deformations of the theory.  They induce a classical RG flow, which can 
lead to an IR fixed point, with the isotropic $z=1$ scaling in the deep 
infrared regime.  As usual in effective field theory, terms of higher order 
in derivatives, or involving additional time derivatives, are of higher 
dimension and therefore superficially irrelevant around the UV fixed point.  

Compared to general relativity, the minimal model is different in three 
interconnected ways:  It has one fewer gauge symmetry per spacetime point, 
its field multiplet has one fewer field component per spacetime point (since 
$N$ is independent of $x^i$), and it propagates an additional scalar graviton 
polarization in addition to the standard tensor polarizations, at least around 
flat spacetime.  While the number of gauge symmetries and field components may 
not be observable, the number of propagating graviton polarizations is.  

\subsection{The nonprojectable case}

Another possibility is to insist on matching the field content of general 
relativity, and promote the lapse $N$ to a spacetime field.  This is the 
``nonprojectable'' theory \cite{mqc,lif}.  If we postulate the same 
$\diff(M,\CF)$ gauge symmetry as in the projectable case, the generic 
action will contain new terms, constructed from the new ingredient 
$a_i\equiv\p_iN/N$.  The general theory with such new terms is sometimes 
referred to in the literature as the ``healthy extension'' of the projectable 
theory.  This is a misnomer -- indeed, the basic rules of effective field 
theory clearly instruct us to include all terms compatible with the postulated 
symmetries, since such terms would otherwise be generated by quantum 
corrections.  Hence, including all terms compatible with the gauge symmetry 
should not be called a ``healthy extension'' of the nonprojectable theory; 
it is just the correct implementation of the assumptions of the nonprojectable 
theory.  (For a recent review of the nonprojectable theory, see \cite{bpsgbh}.)

In contrast, leaving the $a_i$-dependent terms artificially out deserves to be 
called an ``unhealthy reduction'' of the projectable theory.  Such an unhealty 
reduction could only be justified if it is protected by additional symmetries.  
However, a closer analysis of the unhealthy reduction indeed reveals 
difficulties with the closure of the constraint algebra and no new gauge 
symmetry \cite{emil,henx}, possibly with the interesting exception of the deep 
infrared limit \cite{rest}.  

The nonprojectable model may be described in terms of the same field 
content as general relativity, but the scalar graviton polarization is 
still present in its physical spectrum.  In Part~2, we discuss a mechanism 
proposed in \cite{gen}, which eliminates the scalar graviton, by enlarging 
the gauge symmetry to ``nonrelativistic general covariance.'' 

\subsection{Entropic origin of gravity?}

There is another concept originally introduced in \cite{mqc,lif} which has 
caused some level of confusion in the literature: the ``detailed balance'' 
condition.  This concept has its roots in nonequlibrium statistical mechanics 
and dynamical critical phenomena.  Oversimplifying slightly, the theory is 
said to be in detailed balance if the potential in (\ref{smint}) is of a 
special form, effectively a square of the equations of motion associated with 
a (Euclidean-signature) theory in $D$ dimensions with some action $W$.  For 
example, the Lifshitz scalar (\ref{slifs}) is in detailed balance, with 
$W=\frac{1}{2}\int d^D\bx\,\p_i\phi\p_i\phi$.  

In \cite{mqc,lif}, this condition was suggested simply as a technical trick, 
which can possibly reduce the number of independent couplings in $\CV$, 
if one can show that detailed balance is preserved under renormalization 
(which is the case in many nongravitational examples in condensed matter).  
If one is interested in getting close to general relativity with a small 
cosmological constant at long distances, detailed balance would clearly 
have to be broken, at least in the minimal theory.  If that breaking happens 
only at the level of relevant deformations, the restrictive power of the 
detailed balance condition can still be useful for constraining the terms 
whose dimension equals that of the kinetic term.  

Is it possible that the detailed balance condition could play a more physical 
role in our understanding of gravity?  While this question remains open, one 
intriguing analogy seems worth pointing out:  When gravity with anisotropic 
scaling satisfies the detailed balance condition, its path integral 
in imaginary time is formally analogous to the Onsager-Machlup theory of 
nonequilibrium thermodynamics \cite{om1,om2}.  In this path-integral 
formulation of nonequilibrium systems, a collection of thermodynamic variables 
$\Phi^a$ is governed by the Onsager-Machlup action, given -- up to surface 
terms -- by
\be
S=\frac{1}{2}\int dt\,d^D\bx\left\{\dot\Phi^aL_{ab}\dot\Phi^b+
\frac{\delta W}{\delta\Phi^a}L^{ab}\frac{\delta W}{\delta\Phi^b}\right\}.  
\ee   
Here the Onsager kinetic coefficients $L_{ab}$ represent a metric on the 
$\Phi^a$ space, $\frac{\delta W}{\delta\Phi^a}$ are interpreted as entropic 
forces, and the action $W$ itself plays the role of entropy!  

This analogy leads to a natural speculation, implicit in \cite{mqc,lif}, 
that the nature of gravity with anisotropic scaling is somehow entropic.  
It would be interesting to see whether this analogy can be turned into 
a coherent framework in which some of the intriguing recent ideas about the 
entropic origin of gravity \cite{everlinde} (also \cite{jacobson}) and 
cosmology \cite{easson} can be made more precise.  

\subsection{Causal dynamical triangulations and the spectral dimension of 
spacetime}

In the study of quantum field theory, it is often useful to construct the 
system by a lattice regularization, and study the approach to the continuum 
limit using computer simulations.  In the context of quantum gravity, it is 
natural to define the lattice version by summing over random triangulations 
of spacetime.  This approach works well in spacetimes of two Euclidean 
dimensions, where the system can be solved exactly in terms of matrix models. 
However, extending this success to higher dimensions has proven frustratingly 
difficult, with random triangulations typically yielding branched polymers or 
other phases with fractional numbers of macroscopic dimensions in the 
continuum limit.  

In the past few years, a major breakthrough on this front has begun to emerge 
in the causal dynamical triangulations (CDT) approach to lattice gravity (see 
\cite{ajlre} for a review).  In the CDT approach, the pathological continuum 
phases are avoided by changing the lattice rules slightly:  The random 
triangulations that contribute to the partition sum are constrained to respect 
a preferred foliation of spacetime by fixed (imaginary) time slices.  This 
seemingly innocuous change of the rules turns out to be relevant, in the 
technical RG sense:  It leads to a different continuum limit, with much more 
attractive physical properties.  The macroscopic dimension of spacetime in 
this continuum limit appears to be four, as is indicated by the measurement of 
the so-called ``spectral dimension'' $d_s$ of spacetime \cite{ajl} at the long 
distance limit in the lattice simulation (with $d_s= 4.02\pm 0.1$ reported in 
\cite{ajl}).  This is a promising and exciting result, suggesting that 
perhaps for the first time, we might be close to sensible continuum results 
in lattice quantum gravity! 

Clearly, the relevant change of the rules that makes all the difference in 
the lattice implementation, namely that spacetime is equipped with a preferred 
foliation structure, is very similar to the starting point of the analytic 
approach to quantum gravity with anisotropic scaling.  It is natural 
to conjecture that {\it the CDT formulation of quantum gravity 
represents a lattice version of Lifshitz gravity with anisotropic scaling}.  

The first nontrivial piece of evidence for this conjecture was presented in 
\cite{spdim}.  One of the surprises of \cite{ajl} was not only that
 $d_s\approx 4$ at long distances, but also that at shorter distances, before 
the lattice artifacts kick in, $d_s$ undergoes a smooth crossover to 
$d_s\approx 2$.  How can the effective dimension of spacetime change 
continuously from four at long distances to two at short distances?  An 
analytic explanation was offered in \cite{spdim}:  The spectral dimension is 
a precisely defined geometric quantity, and it can be calculated 
systematically in the continuum approach to quantum gravity with anisotropic 
scaling.  In the mean-field approximation around the flat spacetime, the 
result is \cite{spdim} 
\be
\label{spd}
d_s=1+\frac{D}{z}.
\ee
Hence, if the gravity theory flows from a $z=3$ UV fixed point to a $z=1$ 
IR fixed point, the qualitative crossover of $d_s$ observed in \cite{ajl} 
is reproduced.  The topological dimension of spacetime is always four, 
but the {\it spectral\/} dimension changes because of the anisotropic scaling 
at short distances.  

This argument can be turned around, leading to a prediction:  For example, 
in $2+1$ dimensions (not studied in \cite{ajl}), the value of $z$ required at 
the UV fixed point for power-counting renormalizability and UV completeness 
is $z=2$, while the theory still flows to $z=1$ in the IR.  The Lifshitz 
gravity formula (\ref{spd}) then predicts that the CDT formulation of $2+1$ 
gravity should find a crossover from $d_s=3$ at long distances to $d_s=2$ at 
short distances.  This prediction was beautifully confirmed in the CDT lattice 
approach in \cite{benedetti}.  

\subsection{Phases of gravity}

Additional evidence for the conjecture relating CDT lattice gravity and 
the continuum gravity with anisotropic scaling comes from comparing the phase 
diagrams of the two approaches.  

Recall first the phase structure of the Lifshitz scalar,  first investigated 
in \cite{michelson}.  Including the relevant deformations, and a $\phi^4$ 
self-interaction for stabilization, the theory is given by
\be
\label{sldef}
S=\frac{1}{2}\int dt\,d^D\bx\left\{\dot\phi^2-(\Delta\phi)^2
-\mu^2\p_i\phi\p_i\phi-m^4\phi^2-\lambda\phi^4\right\}.
\ee
With $\lambda>0$, depending on the values of $m^4$ and $\mu^2$, the theory 
can be in three phases.  At positive $\mu^2$, we obtain the standard 
disordered and and uniformly ordered phase, in which the vacuum expectation 
value of $\phi$ either vanishes or takes a constant nonzero value.  At 
negative $\mu^2$, a new, spatially modulated phase appears: In this phase, the 
vacuum condensate of $\phi$ is a periodic function along a spontaneously 
chosen spatial direction.  The phase transition lines meet in the tricritical 
$z=2$ point at $\mu=m=0$.  In the mean field approximation, the three phase 
transition lines joining at the tricritical point share a common tangent 
there \cite{michelson}; this feature is erased by quantum corrections, 
and for tricritical Lifshitz points with multi-component order parameters.  

\begin{figure}[tbp]
\centering
\includegraphics[width=2in]{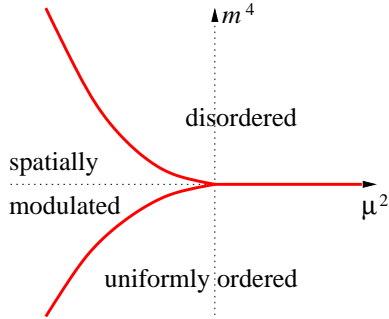}
\caption{The mean-field phase diagram of the Lifshitz scalar theory 
(\ref{sldef}).}
\label{fig0}
\end{figure}

In the CDT approach to quantum gravity in $3+1$ dimensions, three phases have 
also been observed, referred to as A, B and C \cite{ajlre}.  One appears to 
give rise to a macroscopic de~Sitter-like universe, while the other two 
attracted less attention at first, until recently \cite{ambjornhl}.  The lines 
of phase transitions between these phases meet at a tricritical point, whose 
properties have not been explored in much detail on the lattice yet.  

The phase diagram of gravity with anisotropic scaling \cite{hmtz} exhibits 
the same qualitative structure, with several phases organized around a 
multicritical point (see \cite{hmtz} for details).  For simplicity, we 
illustrate this by considering the case of the projectable theory in $2+1$ 
dimensions, where the generic power-counting renormalizable potential is 
\be
\CV=\alpha R^2-\beta R +\gamma.
\ee
Up to a sign, the value of $\alpha$ can be absorbed into a rescaling of 
space versus time.  The remaining sign determines whether we are in real 
or imaginary time.  The terms with the $\beta$ and $\gamma$ couplings, which 
roughly play the role of the (inverse) Newton constant and the cosmological 
constant, represent relevant deformations.  In the mean-field approximation, 
the phases are classified by assuming the FRW ansatz for the metric (with the 
spatial slices compact spatial slices $\Sigma=S^2$, as in CDT), and finding 
the vacuum solutions by solving the Friedmann equation.  It turns out that -- 
as in the Lifshitz scalar -- there are three phases, which meet at the tritical 
$z=2$ point with $\beta=\gamma=0$.  Amusingly, the phase transition lines also 
share a common tangent at the tricritical point, just as in the case of 
the Lifshitz scalar.  We again expect that quantum corrections will modify 
this behavior, without changing the qualitative structure of the phase 
diagram.  

\begin{figure}[tbp]
\centering
\includegraphics[width=2in]{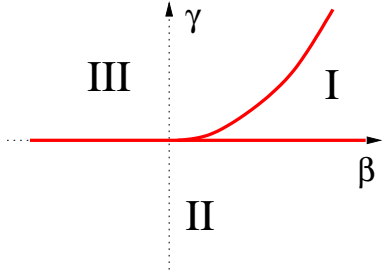}
\caption{The mean-field phase diagram of gravity with anisotropic scaling in 
$2+1$ dimensions with the compact spatial slices $\Sigma=S^2$.}
\label{fig2}
\end{figure}

The nature of the three phases can be analyzed in real or imaginary time.  
In real time, Phase~I corresponds to a global de~Sitter-like spacetime, 
Phase~II describes a recollapsing cosmology with a big bang and a big 
crunch, while Phase~III breaks time reversal spontaneously, with an 
expanding big-bang cosmology or a contracting cosmology with a big crunch.  
In imaginary time, Phase~I yields a compact geometry on $S^3$ much like the 
shape found in \cite{benedetti}, Phase~II is a Euclidean bounce, and in 
Phase~III, there are no solutions satisfying our maximally symmetric FRW 
ansatz.

This similarity between the phase diagram of quantum gravity with anisotropic 
scaling and the phase diagram found in the CDT approach represents further 
evidence \cite{hmtz} for our conjecture that these two approaches to quantum 
gravity are intimately related.  Another universal lesson emerging from our 
analysis of the phase structure of quantum gravity with anisotropic scaling in 
\cite{hmtz} is that spatially modulated phases of gravity should be possible.  

\section{General Covariance in Gravity with Anisotropic Scaling}

In order to eliminate the extra scalar polarization of the graviton, gravity 
with anisotropic scaling which enjoys an extended gauge invariance was 
proposed in \cite{gen}.  The gauge symmetry in question is an extension of 
the foliation-preserving diffeomorphisms by an Abelian gauge symmetry, and 
can be interpreted as a nonrelativistic form of general covariance.  The 
number of independent symmetries per spacetime point is the same as in 
general relativity, with the Abelian symmetry playing the role of linearized 
spacetime-dependent time reparametrizations.  This extended symmetry preserves 
the preferred spacetime foliation and the privileged role of time, but 
eliminates the scalar polarization of the graviton.  

\subsection{Fields and symmetries}

We start with the minimal projectable theory reviewed in Part~1.  It was 
noticed in \cite{lif} that at $\lambda=1$, this theory exhibits in 
the linearized approximation around flat spacetime an enhanced symmetry, 
which acts only on the shift vector, 
\be
\label{alphax}
\delta N_i=\p_i\alpha,
\ee
with $\alpha(\bx)$ a time-independent local symmetry generator.  Promoting 
this symmetry to a spacetime-dependent gauge symmetry of the full nonlinear 
theory will lead to our desired nonrelativistic general covariance.  

Extending (\ref{alphax}) to a gauge symmetry requires new fields beyond 
the minimal gravity multiplet $g_{ij}$, $N_i$ and $N(t)$.  Already at the 
linearized level, we need to introduce a new field $A$ which transforms under 
$\alpha(\bx,t)$ as the time component of an Abelian gauge field.  In the 
interacting theory, this transformation rule becomes
\be
\delta A=\dot\alpha -N^i\p_i\alpha.
\ee
The new field $A$, and the new gauge symmetry $\alpha$, have an elegant and 
geometric interpretation \cite{gen} in the context of a nonrelativistic $1/c$ 
expansion of relativistic gravity: $A$ is simply the subleading term in the 
$1/c$ expansion of the relativistic lapse function, and $\alpha$ is the 
subleading, linearized part of spacetime-dependent time reparametrizations. 

Unfortunately, in dimensions greater than $D=2$, this is not the whole story.  
When $D>2$, the linearized symmetry (\ref{alphax}) does not extend to a 
symmetry of the interacting, nonlinear theory, and therefore cannot be 
straightforwardly gauged.  In order to fix this obstruction, a new field 
$\nu$ was introduced in \cite{gen}.  This ``Newton prepotential'' transforms 
as a Goldstone field,
\be
\delta\nu=\alpha.
\ee
The introduction of the Newton prepotential allows (\ref{alphax}) to be 
extended to a symmetry of the nonlinear theory, which can then be gauged 
by the standard coupling to the gauge field $A$.  Unlike the rest of the 
gravity multiplet, the Newton prepotential does not appear to have a natural 
geometric interpretation in terms of the $1/c$ expansion in the metric 
formulation of relativistic gravity.

\subsection{The Lagrangian and Hamiltonian formulations}

The systematic construction of an action invariant under the extended gauge 
symmetries leads to the following minimal theory \cite{gen}, 
\be
\label{fullact}
S=\frac{2}{\kappa^2}\int dt\,d^D\bx\,\sqrt{g}\left\{N\left[K_{ij}K^{ij}-K^2-\CV
+\nu\,\Theta^{ij}\vphantom{g^{ik}}\left(2K_{ij}+\nabla_i\nabla_j\nu\right)
\right]-A\,(R-2\Omega)\right\}.  
\ee
Here $\Theta^{ij}$ a short-hand for $\Theta^{ij}=R^{ij}-\frac{1}{2}g^{ij}R+
\Omega g^{ij}$, and $\Omega$ is a new relevant coupling constant, of the same 
dimension as the cosmological constant $\Lambda$.  It controls the scalar 
curvature of the spatial slices in the preferred foliation $\CF$ of spacetime, 
and it makes sense to refer to $\Omega$ as the ``second cosmological 
constant.''   The form of the potential $\CV$ is again unconstrained by the 
symmetries, just as in the minimal theory.  

The theory can also be rewritten in the Hamiltonian formalism \cite{gen}, which 
offers a more systematic way for studying the gauge symmetry structure and 
counting the number of propagating degrees of freedom without having to 
resort to sometimes unreliable linearizations around a chosen background.  
The Hamiltonian constraint algebra exhibits an intriguing mixture of first- 
and second-class constraints, and confirms that the theory propagates only the 
tensor graviton polarizations.  The scalar graviton mode is seen as a gauge 
artifact of nonrelativistic general covariance.

\subsection{Comparing to general relativity in the infrared}

Since the spectrum of propagating gravitons -- and gravitational waves -- in 
the long-distance limit of our gravity with nonrelativistic general covariance 
matches that of general relativity, it is natural to extend this comparison to 
the long-distance limits of the full nonlinear theories.   

Some first steps in this direction were made in \cite{gen}.  First, a 
simple conceptual argument implies that the Schwarzschild spacetime is an 
exact solution of the infrared limit of our theory.  This bodes well for 
the standard tests, since it suggests that in the infrared regime, the $\beta$ 
and $\gamma$ parameters of the PPN formalism take their relativistic value, 
equal to one.  

The equation of motion associated with the variation of $A$ constrains the 
spatial scalar curvature to be constant, $R=2\Omega$.  At first, it might 
seem that this equation might be difficult to reconcile with the existence of 
interesting cosmological solutions.  However, this issue can be avoided in 
several different ways, and interesting cosmological solutions can be found 
\cite{gen}.  In fact, the theory has a phenomenologically attractive feature:  
it seems to prefer cosmologies whose preferred spatial slices are flat.  

Perhaps the biggest challenge for this program is to explain why the infrared 
limit should exhibit Lorentz invariance, to the high level of accuracy 
required by observations.  While the theory may naturally flow to $z=1$ 
at long distances, different species of low-energy probes may experience 
distinct effective limiting speeds of propagation, not equal to the speed 
of light.  Setting all these speeds equal to $c$ would represents a rather 
unpleasant amount of fine tuning.  While this problem remains unsolved in the 
theory with nonrelativistic general covariance as well, it is intriguing that 
-- unlike in the minimal theory -- global Lorentz symmetries of the flat 
spacetime can be embedded into the extended gauge symmetry of our generally 
covariant theory \cite{gen}.  

\section{Conclusions}

If one's agenda is to construct a theory with anisotropic scaling which 
resembles general relativity in the infrared, the generally covariant model 
of \cite{gen} appears to be a step in the right direction, since its extended 
gauge symmetry eliminates the scalar graviton from the theory, leaving only 
the physical tensor polarizations.  The resulting infared limit has the 
Schwarszchild geometry as an exact solution, suggesting that the theory is 
likely compatible with the standard solar-system tests.  

The price paid is the introduction of the rather mysterious Newton 
prepotential $\nu$ in \cite{gen}.  This field does not appear to have a clear 
geometric interpretation in the $1/c$ expansion of the standard metric 
formulation of relativistic gravity.  Moreover, the introduction of $\nu$ 
leads to a new proliferation of gauge invariant terms that can appear in the 
action, both for pure gravity and in its coupling to matter 
\cite{patrick,dasilva}.  Clearly, a better understanding of the role of the 
Newton prepotential is desirable before one can seriously discuss detailed 
phenomenological constraints on models of gravity with nonrelativistic general 
covariance.   

\acknowledgments
It is a pleasure to thank Kevin Grosvenor, Charles Melby-Thompson, Cenke Xu 
and Patrick Zulkowski for enjoyable collaborations on the topics discussed 
in this paper.  This paper is based on the invited talks delivered at the 
{\it GR 19 Conference}, Ciudad de M\'exico, July 2010.  I wish to thank Cedric 
Defayett, Fay Dowker, Don Marolf and Shiraz Minwalla for their invitation and 
hospitality.  This work has been supported by NSF Grant PHY-0855653, by DOE 
Grant DE-AC02-05CH11231, and by the BCTP.  

\bibliographystyle{JHEP}
\bibliography{grx}
\end{document}